# Many-body effects in a quasi-one-dimensional electron gas

Sanjeev Kumar[1,3*], Kalarikad J. Thomas[1,3], Luke W. Smith[2], Michael Pepper[1,3], Graham L. Creeth[1,3], Ian Farrer[2], David Ritchie[2], Geraint Jones[2] and Jonathan Griffiths[2]

[1]*London Centre for Nanotechnology, 17-19 Gordon Street, London WC1H 0AH, UK*

[2]*Cavendish Laboratory, J.J. Thomson Avenue, Cambridge CB3 OHE, UK*

[3]*Department of Electronic and Electrical Engineering, University College London, Torrington Place, London WC1E 7JE, UK*

*Correspondence to: sanjeev.kumar@ucl.ac.uk

**Abstract:** We have investigated electron transport in a quasi-one dimensional (quasi-1D) electron gas as a function of the confinement potential. At a particular potential configuration, and electron concentration, the ground state of a 1D quantum wire splits into two rows to form an incipient Wigner lattice. It was found that application of a transverse magnetic field can transform a double-row electron configuration into a single-row due to magnetic enhancement of the confinement potential. The movements of the energy levels have been monitored under varying conditions of confinement potential and in-plane magnetic field. It is also shown that when the confinement is weak, electron occupation drives a reordering of the levels such that the normal ground state passes through the higher levels. The results show that the levels can be manipulated by utilising their different dependence on spatial confinement and electron concentration, thus enhancing the understanding of many body interactions in mesoscopic 1D quantum wires.

There have been many studies of a strongly confined, non-interacting one-dimensional (1D) electron or hole gas in which the electrostatic confinement is provided by a voltage applied to split or patterned gates [1-4]. In general the results have been in agreement with the predictions of established theories, such as Luttinger theory and spin-charge separation [5], with the exceptions being the 0.7 anomaly arising from a partial spin polarisation [6], and a spin polarisation when the momentum degeneracy in the ground 1D subband is lifted by a source-drain voltage [7].

Considerable alteration of the electronic properties can occur with modification of the confinement potential. For example, by inducing a reflection at the channel exit the system can display Kondo behaviour which is additive to the 0.7 anomaly, a many body effect [8]. Recently it has been found that when the confinement is weak the ground state conductance is no longer $2e^2/h$ but $4e^2/h$ [9-11]. This transition was attributed to the splitting of electrons occupying the first subband into two separate rows, each with a conductance of $2e^2/h$; these rows being the lowest energy state minimising the electron-electron interaction as the confining harmonic oscillator potential weakens. Such an "Incipient Wigner Lattice" was predicted by theories suggesting that the first stage in such a formation was the distortion of the line of electrons into a "zig-zag" configuration, which on increasing (decreasing), the repulsion (confinement) resulted in separation into two rows [12-15]. A similar effect has been calculated for two interacting particles in a square box [16]. Although a zig-zag cannot be determined by measurement of conductance behaviour it was possible to ascertain that the first stage in the breakdown of the spatially quantised ground state was the formation of a "bonding" state followed by a separation into two rows [10]. In the regime where electron-electron interaction is paramount, a number of 1D states are predicted including a ferromagnetic state [17] and various spin phases, which has aroused a considerable interest in the research of interacting fermions in quantum wires, such as Ref. [18].

In this Letter we present results showing in detail the competing roles of spatial confinement and the electron-electron interaction on the 1D quantisation. We show how it is possible to alter the plateau sequence by a modification of the higher levels as well as the ground state. A surprising change in the nature of the ground state is shown to occur. In addition we discuss the effects of a transverse magnetic field which is known to enhance confinement and lead to level depopulation.

The device used in the present work was fabricated from a delta doped GaAs/AlGaAs heterostructure grown using a molecular beam epitaxy (MBE), where a 2DEG formed 300 nm beneath the interface had a mobility in the dark (light) of $1.85 \times 10^6$ cm$^2$/Vs ($3.1 \times 10^6$ cm$^2$/Vs) and electron density of $9.9 \times 10^{10}$ cm$^{-2}$ ($2.0 \times 10^{11}$ cm$^{-2}$). A pair of split gates of length 0.4 μm, and width 0.7 μm and a top gate of length 1$\mu$m separated by 200 nm thick insulating layer of cross-linked PMMA were patterned by a standard lithographic technique [9]. Two-terminal differential conductance (*G*) measurements were performed using an excitation voltage of 10 $\mu$V at 73 Hz in a cryofree dilution refrigerator with electron temperature of 70 mK.

The inset of Fig. 1(a) is a schematic diagram of a typical device used in this work with a pair of split gates and a uniform top gate [19]. Figure 1(a) shows the conductance characteristics of the device as a function of split gate voltage ($V_{sg}$) for a constant top gate voltage ($V_{tg}$). Strong (weak) confinement is on the left (right). Weakening the confinement results in the loss of the $2e^2/h$ plateau and $4e^2/h$ becomes the ground state which is indicated by the 'blue' trace at $V_{tg}$=-3.25 V; eventually $2e^2/h$ reappears when the confinement is further weakened. Figure 1(b) is the greyscale plot of transconductance ($dG/dV_{sg}$) as a function of $V_{sg}$ and $V_{tg}$ for data in (a) showing what appears as an anti-crossing of the subbands around $V_{tg}$=-3.25 V corresponding to double row formation. Inset shows a magnified view of the anti-crossing.

The measured conductance depends on the number of occupied subbands below the Fermi level. The first plateau, arising from the ground state, corresponds to a transverse wavefunction $\Psi_0$, termed the 0-state, which figuratively has a half-sine like characteristic. The second plateau of value $4e^2/h$ resulting from occupation of the second level, is represented by the transverse wavefunction, $\Psi_1$, which is sine-like [Fig. 1(c)], and is termed the 1-state. As the confinement is weakened, $\Psi_0$ and $\Psi_1$ are no longer the simple single-electron wavefunctions but are modified by the electron-electron interaction . A detailed picture is obtained by considering the greyscale plot [Fig. 1(b)] which shows the variation in the energy levels as the confinement is altered, the left, (right) is the strongest, (weakest), confinement. A significant feature is the narrowing of the difference between the 0- and 1-states due to more rapid drop of the 1-state which eventually converges with the 0-state. We note that the 1-state moves roughly parallel with the higher energy states. One possible explanation of this behaviour is the different role of the electron repulsion in the 0- and 1-state as the confinement weakens. In the 1-state the electrons can describe a highly correlated motion in alternate lobes of the state. The state drops in energy as the confinement width increases which also reduces the electron-electron repulsion. The 0-state has the electrons confined near the middle of the channel and does not experience such a reduction in electron repulsion as the 1-state. Hence the 0-state energy only changes due to the weaker confinement, rather than the interaction, and so reduces at a slower rate than the 1-state. When the levels overlap the result of the mixing is a hybridised wavefunction which is preferentially located at the edges, corresponding to the formation of two rows, producing a conductance of $4e^2/h$ (schematically shown by dotted green and blue traces in Fig. 1(c)). When the confinement potential is further weakened, the 1-state drops below the 0-state and the conductance plateau $2e^2/h$ returns reflecting the reduced role of the interaction in the new ground state.

The device was further characterised by performing dc-bias spectroscopy at $V_{tg}$ =-3.25 V where we see the first plateau appearing at $4e^2/h$. Figure 2(a) shows the dc-bias characteristics of the device from 0 to -3 mV. It is observed that there is the usual $0.25(2e^2/h)$ feature in the strong dc-bias which indicates that either the two rows move together initially or else the source-drain voltage assists one to form before the other. Structure at $0.5(2e^2/h)$ is also observed, this feature is not observed for strong confinement it is indicative of the addition of the conductance from two separate rows [10,11]. Figure 2(e) is the greyscale plot of data in (a) where $0.25(2e^2/h)$ and $0.5(2e^2/h)$ structures can be traced by looking at the dark, black regions which extend almost parallel to each other as the source-drain bias is increased. In addition, there are half integer plateaux due to the lifting of the momentum degeneracy which are symmetric about zero source-drain voltage [20].

Figure 2(b)-(d) show the dc-bias results in the presence of a small transverse magnetic field (perpendicular to 2D electron gas plane), $B_{tr}$ of 0.2, 0.3 and 0.45 T, respectively. With an increase in $B_{tr}$, a feature at $2e^2/h$ starts appearing and at 0.45 T the first plateau at $2e^2/h$ is completely restored. Such a removal of the direct jump to $4e^2/h$ is due to the additional confinement provided by the magnetic field [2]. Broadened plateaux are found at integer multiples of $2e^2/h$ at zero dc-bias, on increasing the dc-bias the usual $0.85(2e^2/h)$ structure appears [21]. Further increase in the dc-bias results in the appearance of the $0.25(2e^2/h)$ structures and the $0.5(2e^2/h)$ plateau is much reduced. The greyscale plots of $dG/dV_{sg}$ for data in Fig. 2(b)-(d) are shown in Fig. 2(f)-(h), respectively which give further insight into the disappearance of the double row. It may be noticed that with an increase in $B_{tr}$, plateaux become sharp and clear due to a reduction in backscattering. Some additional features are observed in (g) and (h) such as the splitting of plateaux at zero source-drain bias near $2e^2/h$ which could be due to the approach to the quantum Hall effect [22].

The device was measured again in a different cool down [23], and the results are shown in Fig 3(a) for zero magnetic field and Fig. 3(b) with a magnetic field, $B_{||}$ of 12 Tesla applied parallel to the electron gas. Figures 3(c) and (d) are the greyscale plots of $dG/dV_{sg}$ for data shown in Figs. 3(a) and (b), respectively. The most significant feature of Fig. 3(a) is that as the confinement weakens the 0-state passes through the 1-state giving rise to an anti-crossing and a jump to $4e^2/h$ as indicated in the 'blue' trace at around $V_{tg}$=-8.53 V in Fig. 3(a) [Also, see Fig. 3(c)]. It may be noted that a number of traces in close proximity to the 'blue' trace exhibited direct jumps to $4e^2/h$ which indicates that the two rows are stable in this regime. We note that as there is only one conductance plateau the two levels formed by the hybridisation are degenerate, i.e. the two rows. The difference in the rate of movement of the two levels illustrates that their energies are not just determined by spatial confinement which would cause them to move in a similar manner. The energy of the 0-state is more insensitive to the confinement and after crossing the 1-state continues to rise relative to the other higher states (Fig. 3(c)). This feature is also indicated by the closer proximity of the conductance plots to each other as a function of split-gate voltage occurring when the 1-state drops to become the new ground state [see, traces on the right of 'green' trace in Fig. 3 (a)]. If the effect was purely electrostatic, due to the dependence of channel formation on the top- and split-gate voltages, then a gradual transition may be expected. The sharpness of the transition indicates that the ground state formed by the 1-state is less dependent on the carrier concentration, and electron-electron interaction, than that when the 0-state was the ground state. As the confinement is further weakened, when $V_{tg} \leq -8.75$ V, the 0-state rises above the 1-state and anti-crosses the 2-state giving a jump to $6e^2/h$ as there is no longer a separate 2-state, as shown by 'green' trace in Fig 3(a). This effect consequently occurs only when the $2e^2/h$ is re-introduced as the first plateau, due to the 1-state being the new ground state. An important aspect of this effect is that jumps due to the omission of higher plateaux are only observed when the 0-state is passing up through the levels and consequently the first $2e^2/h$ is present. The return of the $4e^2/h$ plateau is accompanied by a tendency for the $6e^2/h$ to disappear with a jump from $4e^2/h$ to $8e^2/h$, however the greyscale plot is not sufficiently clear for the cause to be established unambiguously.

The role of the 12 Tesla in-plane magnetic field is immediately apparent by the overall increased closeness of the conductance plots in Fig. 3(b) arising from the lifting of the spin degeneracy. This is clearest for strong confinement where the conductance is given by $Ne^2/h$, where N is 1,2,3,5.. The missing 4 and 6 arise from overlap of spin down and spin up levels from different subbands [see first 'black' trace in Fig 3(b)]. Assuming the spin down ↓ state to be the lowest spin energy, and adopting the same indexation of states as previously, we see from inspection of Fig. 3(b) that a merger of the 1↑ and 2↓ states occurs and that as the channel is widened they stay together. In a similar vein as the channel widens the overlap of the 0↑ and 1↓ states takes place and in the conductance plot the plateau with integer value N=2 disappears ('dark-black' trace near $V_{sg}$=-3.4 V in Fig. 3(b)). As the channel widens they do not diverge but the levels have formed a stable state, this may be related to previous observations of energy level locking in parallel quantum wires [24,25]. Possibly this effect is enhanced by the opposite spin polarisations of the two levels participating in the locking.

Widening the channel further results in a complex pattern of overlapping levels which is similar to another example of level hybridisation due to quasi 1D channels formed in coupled electron gases [26]. When the $V_{sg}$ threshold is approximately -3.13 V, a gap appears in the traces and an integer value of N=4 appears although N=2 continues to be absent; from then on N=3 is very weak and disappears. The gap in the characteristics, particularly at $V_{sg}/V_{tg}$=-3.13 V/-8.26 V, 'dark-red' trace, is an indicator of an increased density of states as the levels overlap. A spin-polarised state of the double row is at $2e^2/h$ for $V_{sg}/V_{tg}$=-2.95 V/-8.6 V as indicated by 'light-blue' trace. Stability of the system is shown by the observation of a jump from $0.5e^2/h$ to $2.5e^2/h$, indicated by 'dark-blue' trace at $V_{sg}/V_{tg}$=-2.75 V/-8.8 V which is a spin-polarised form of the jump ($2e^2/h$ to $6e^2/h$) observed in the absence of the magnetic field, indicated by the 'green' trace in Fig. 3(a).

Insight into the processes causing such behaviour can be drawn from the greyscale plot in Fig. 3(d) where we see that the ground $0\downarrow$ changes at a much slower rate than the higher levels so cutting across them and giving rise to the pattern of crossing events. This is similar to the zero magnetic field case but greatly enhanced so accounting for the reduced region of row formation where the $e^2/h$ disappears and the first plateau is $2e^2/h$. The increasing hybridisation as levels move together also explains why a new plateau emerges from one plateau then drops in value before settling at the next quantised value [27]. A significant feature of the conductance plots is the disappearance of the $2e^2/h$ plateau and the reintroduction of $e^2/h$ as the ground state, this arises from the $2\downarrow$ dropping through many states to become the new ground state. This is a surprising manifestation of the modification of the levels by the electron interaction. The 0-state passes through several levels during its trajectory but the results are not sufficiently clear to determine the spin dependence of hybridisation.

In conclusion, we have studied electron transport in a quantum wire where the confinement is altered by a combination of split-gates and a top gate. When the confinement is weakened, conductance plateau at $2e^2/h$ disappears, whereas $4e^2/h$ plateau persists and becomes the new ground state comprising two rows. Monitoring the movement of energy levels, it appears that the electrons in the ground state (0-state) hybridize with the first excited state (1-state) to give rise to $4e^2/h$ as a first plateau, and then further reduction of the confinement results in the 1-state falling below the 0-state, and a $2e^2/h$ plateau re-appears. It has been found that the electron-electron interaction affects the ground state more significantly than the higher levels causing it to move through them as the confinement weakens, so modifying the sequence of quantised plateaux. Such movement of energy levels as a consequence of Coulomb interaction gives further insight on the conductance behaviour of weakly confined quasi-1D systems.

We thank Dr. J. T. Nicholls and Professor G. Gumbs for many fruitful discussions. This work was supported by the Engineering and Physical Sciences Research Council (EPSRC), UK.

**Figure captions**

FIG. 1. (a) Conductance characteristics of the device for various $V_{tg}$ (0 to -3.9 V). Inset shows schematic diagram of a top-gated split-gate device; 'red' are the split gates (SG) and 'orange' is the top gate (TG), and PMMA is shown by a 'grey' layer. (b) Greyscale plot of $dG/dV_{sg}$ for data in (a); here red colour indicates the riser in conductance and the dark regions are the conductance plateaux; (c) Schematic diagram showing the hybridization of ground state and first excited state wavefunction in a double row system. The simple wavefunctions are shown whereas in reality they are heavily distorted by the electron-electron interaction.

FIG. 2. Dc-bias characteristics measured in the presence $B_{tr}$. (a)-(d) show plots of conductance characteristics for source-drain bias of 0 to -3 mV in the presence of $B_{tr}$ of 0, 0.2, 0.3, and 0.45 T, respectively. The traces have been offset horizontally for clarity. The trace on the extreme left (right) corresponds to 0 V (-3 mV) dc-bias, and the successive traces are taken at intervals of 0.5 mV. (e)-(h) represent greyscale plot of $dG/dV_{sg}$ for data in (a)-(d), respectively, red regions are the conductance risers and black regions are conductance plateaux.

FIG. 3. (a) Conductance characteristics of the device in a different cool down for different $V_{tg}$ (-7.5 V to -9.2 V). As the confinement weakens the plateaux corresponding to $2e^2/h$ and $4e^2/h$ smear out and then re-appear. (b) Conductance characteristics in the presence of $B_{ll}$ of 12 T showing Zeeman splitting and crossing of the 0-state with the higher subbands as the confinement weakens. (c) and (d) show $dG/dV_{sg}$ plot represented as greyscale for data in (a) and (b), respectively where dark regions represent the conductance risers and the red regions are the conductance plateaux. Inset of (c) and (d), indicated by arrows, show the magnified view of anti-crossing and crossing, respectively of the 0-state with the 1 and other higher states.

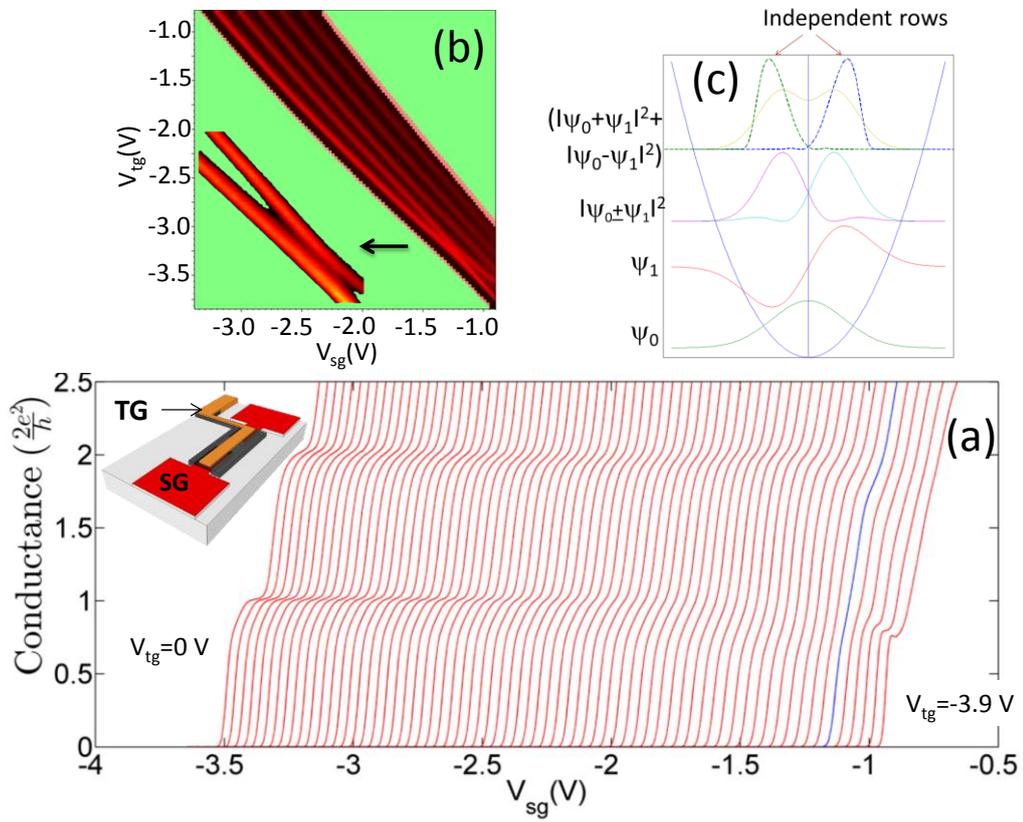

**Figure 1**

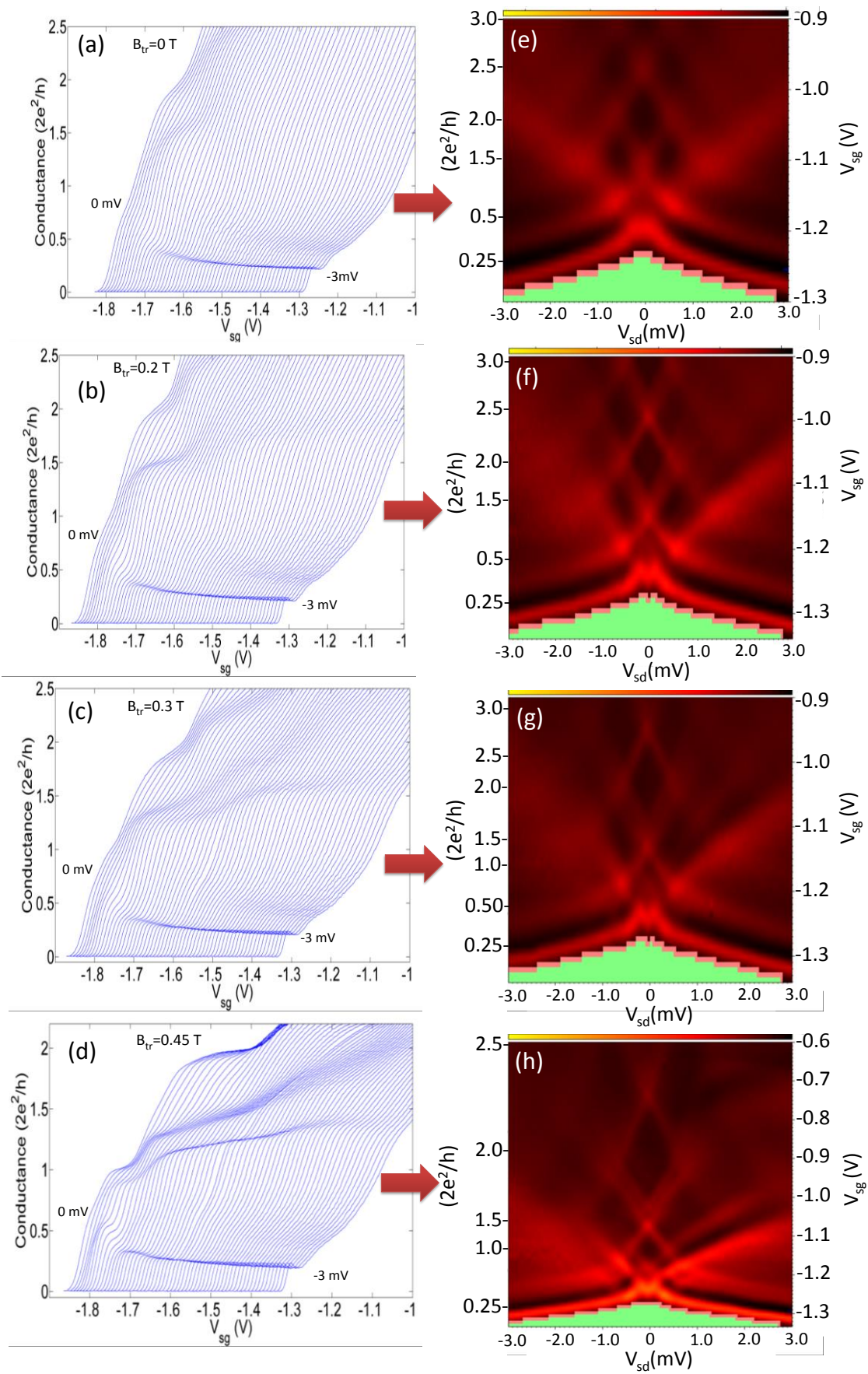

**Figure 2**

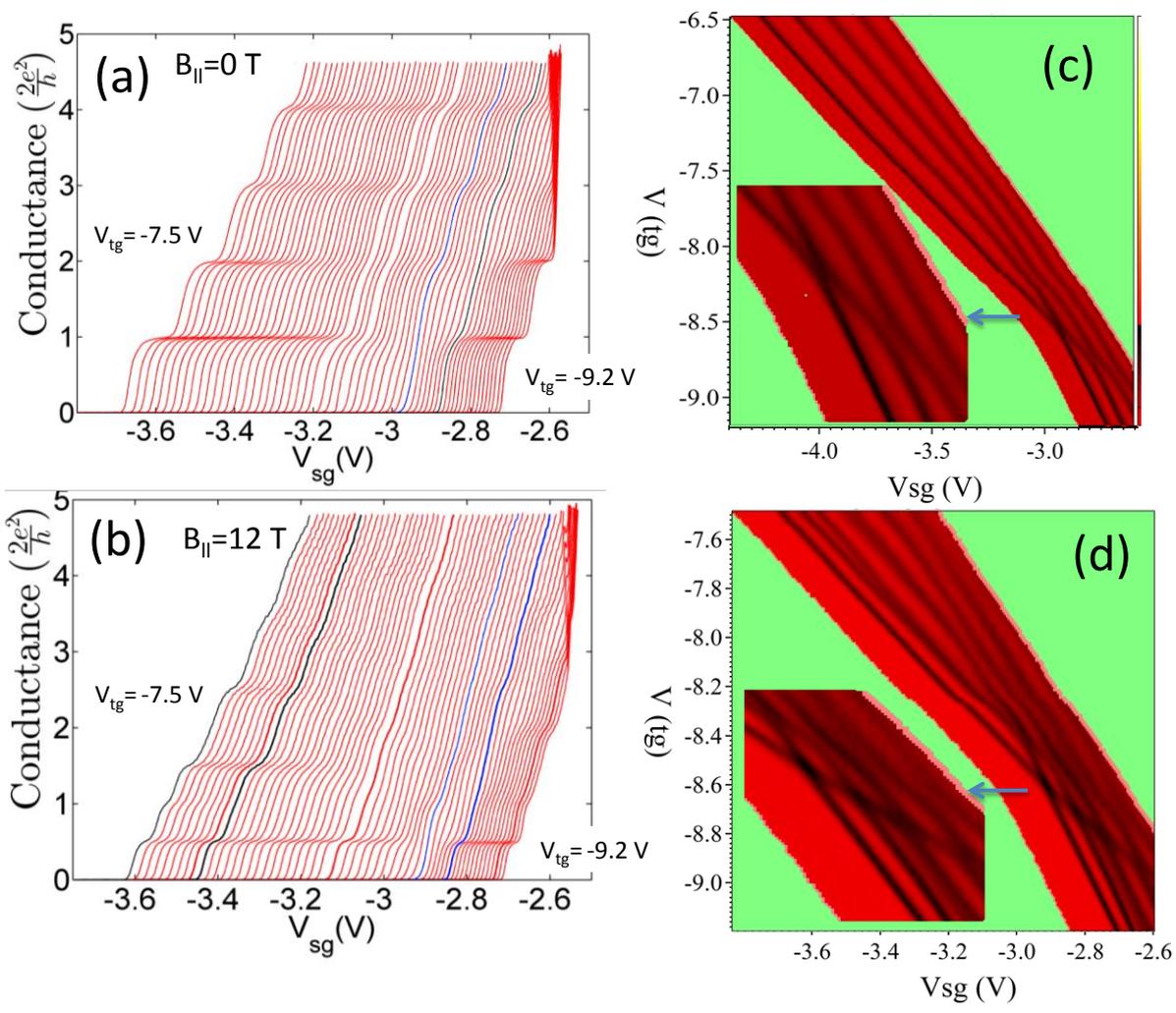

**Figure 3**